\documentclass[twocolumn, aps,showpacs]{revtex4}
\usepackage{graphicx}

\begin{document}
\title{Weakly nonlinear vertical dust grain oscillations \\
in dusty plasma crystals
 \\  in the presence of a magnetic field
 \footnote{Preprint; submitted to \it{Physics of Plasmas}.
}
 }
\author{I. Kourakis}
\altaffiliation[On leave from: ]{U.L.B. - Universit\'e Libre de
Bruxelles, C. P. 231 Physique Statistique et Plasmas,
Association Euratom -- Etat Belge, Boulevard du Triomphe,
B-1050 Brussels, Belgium; also at:
U.L.B., Facult\'e des Sciences Apliqu\'ees - C.P. 165/81
Physique G\'en\'erale, Avenue F. D. Roosevelt 49, B-1050 Brussels,
Belgium}

\email{ioannis@tp4.rub.de}
\author{P. K. Shukla}
\email{ps@tp4.rub.de} \affiliation{Institut f\"ur Theoretische
Physik IV, Fakult\"at f\"ur Physik und Astronomie,
Ruhr--Universit\"at Bochum, D-44780 Bochum, Germany}
\date{Received 06 February 2004; revised 20 April}

\begin{abstract}
The weakly nonlinear regime of transverse paramagnetic dust grain
oscillations in dusty (complex) plasma crystals is discussed.  The
nonlinearity, which is related to the sheath electric/magnetic
field(s) and to the inter--grain (electrostatic/magnetic dipole)
interactions, is shown to lead to the generation of phase
harmonics and, in the case of propagating transverse dust-lattice
modes, to the modulational instability of the carrier wave due to
self--interaction. The stability profile depends explicitly on the
form of the electric and magnetic fields in the plasma sheath. The
long term evolution of the modulated wave packet, which is
described by a nonlinear Schr\"odinger--type equation (NLSE), may
lead to propagating localized envelope structures whose exact
forms are presented and discussed. Explicit suggestions for
experimental investigations are put forward.
\end{abstract}

\maketitle

\section{Introduction}

Dusty (or complex) plasmas (DP) have been attracting an increasing
interest among plasma physicists for more than a decade due to
to the appearance of many novel phenomena.
Of particular importance is the possibility of the existence of new
plasma configurations (states), due to the strong intergrain coupling,
including  the spontaneous formation of crystal--like
DP structures, when charged microparticles (dust grains) are
trapped in the sheath region between the electrodes, in plasma discharge
experiments \cite{PKSbook, Morfill}.
Such crystals, which are generally horizontally arranged in a
position of equilibrium which is levitated above the negative electrode
in discharge experiments, have been shown to support longitudinal (acoustic) as well
as transverse (optic-like) oscillation modes \cite{PKSbook}.

Force equilibrium in DP crystals is "traditionally" ensured by the sheath
electric force which exactly balances gravity at the levitation height.
Crystalline complex plasma structures have been observed in recent rf discharge
experiments \cite{SamsonovNJP} in which the plasma sheath was embedded
in an external magnetic field. Gravity compensation was thus attributed to
magnetic forces, thanks to the paramagnetic properties of the magnetized dust grains.
Theoretical studies then followed for the investigation of conditions
for magnetic-field-assisted crystal equilibria involving paramagnetic charged
dust grains.  The role of various forces acting on paramagnetic grains has
been  discussed in Ref. \cite{YaroNJP}, where magnetic forces (due to magnetic
dipole interactions) have been shown to prevail over the (weaker) electric polarization
forces (especially under experimental conditions considered in Ref. \cite{SamsonovNJP},
where magnetic fields as strong as a few thousand Gauss were used).
According to these considerations, the properties of transverse magnetized dust
lattice (TMDL) oscillations were  investigated \cite{Yaro}, where both
isolated grain linear oscillatory modes and propagating linear waves were
discussed.  The dependence of the dynamics of (linear) grain oscillations
on the specific characteristics of the (inhomogeneous) magnetic field was
studied in detail in Ref. \cite{Yaro}, and the possible occurrence of
a linear instability depending on the field profile was pointed out.

It is now established that the investigation of the linear regime
of a dynamical system only unveils part of its dynamical profile.
Nonlinearity may be present in DP crystal dynamics, either due to intergrain
interactions or due to the sheath environment.
In a generic manner, as the oscillation amplitude becomes large,
nonlinearity is first manifested via the generation of wave harmonics leading
to the amplitude modulation of the waves. Increasing displacements even
further, one may come up with the formation of localized structures (solitons),
which propagate and interact with each other in a remarkably stable manner,
thanks to a compensation between nonlinearity and dispersion.
The present study is devoted to an investigation of weakly nonlinear effects, with
respect to magnetized transverse dust-lattice oscillations.
Nonlinearity is explicitly shown to be related to the magnetic and/or electric
field profile, as well as to the intergrain electrostatic and/or magnetic
(dipole-dipole) interactions.  By using a perturbative approach, the generation of
phase harmonics is elucidated and exact expressions are obtained for the (weak)
vertical displacement of the paramagnetic dust particles.
Once these oscillations propagate in the dusty crystal as a transverse wave,
the amplitude is shown to be potentially unstable to external perturbations,
under conditions which depend explicitly on the magnetic field characteristics.
Finally, the possible formation of localized envelope excitations is discussed,
and the dependence of these coherent structure characteristics on the plasma
parameters is pointed out.

\section{Nonlinear single grain oscillations}

Let us consider the vertical motion of a charged dust grain
(mass $M$ and charge $q$),
subject to an external static electric and magnetic field,
$\mathbf{E}$ and $\mathbf{B}$ respectively, both in the vertical
($\sim \hat z$) direction. The vertical displacement
$\delta z = z - z_0$ from the equilibrium position $z_0$ obeys the
equation of motion
\begin{equation}
\ddot{z} \, = \, \frac{F}{M} - g \, - \, \nu \, \dot{z} \, ,
\label{eqmotion1}
\end{equation}
where we have set $z_0 = 0$. The three terms in the right-hand side (rhs)
of Eq. (1) account for the (sum of the) electric and magnetic forces,
viz. $F = F_e \, + \, F_m$,
in the $z-$ direction, the force of gravity, and
the usual (Epstein) damping term, which involves the phenomenological damping
rate $\nu$ due to dust--neutral collisions.

We shall assume a smooth, continuous variation of the (generally inhomogeneous)
field intensities $\mathbf{E}$ and $\mathbf{B}$, as well as the grain charge
$q$ (which may vary due to charging processes \cite{Ivlev2000PRE622739})
near the equilibrium position $z_0 = 0$. Thus, we may develop
\[
E(z) \approx E_0 \, + \, E'_0 \, z \,+ \, \frac{1}{2} E''_0 \, z^2 \, + \, ...
\, ,
\]
\[
B(z) \approx B_0 \, + \, B'_0 \, z \,+ \, \frac{1}{2} B''_0 \, z^2 \, + \, ...
\, ,
\]
and
\[
q(z) \approx q_0 \, + \, q'_0 \, z \,+ \, \frac{1}{2} q''_0 \, z^2 \, + \, ...
\, ,
\]
where the prime denotes differentiation with respect to $z$ and the
subscript `$0$' denotes evaluation at $z = z_0$, viz.
$E_0 = E(z=z_0)$, $E'_0 = d E(z)/d z|_{z=z_0}$ and so forth.
Accordingly, the electric force $F_e = q(z) E(z)$ and the magnetic force
$F_m = - \partial (m B)/\partial z = - 2 \alpha \, B \, \partial B/\partial z$
(where the grain magnetic moment $\mu$ is related to the grain radius
$a$ and permeability $\mu$ via $m = (\mu - 1) a^3 \,B/(\mu + 2) \equiv
\alpha B$ \cite{Jackson}), which are now expressed as
\[
F_e(z) \approx q_0 E_0 \, + \,
(q_0 E'_0 + q'_0 E_0) \, z \qquad \qquad
\]
\[ \qquad \qquad \,+
\, \frac{1}{2} (q_0 E''_0 + 2 q'_0 E'_0 + q''_0 E_0)
\, z^2 \, + \, ...
\, ,
\]
and
\[
F_m(z) \approx - 2 \alpha B_0 B'_0 \, - \,
2 \alpha ({B'_0}^2 + B_0 {B''}_0)\, z \,\qquad \qquad
\]
\[ \qquad \qquad  -
\, \alpha (B_0 B'''_0 + 3 B_0' {B''}_0)
\, z^2 \, + \, ...
\, ,
\]
may be combined into
\[
F_e + F_m = - \frac{\partial \Phi}{\partial z} \, ,
\]
where we have introduced the phenomenological potential $\Phi(z)$
\begin{eqnarray}
\Phi(z) & \approx & \Phi(z_0) \,
+ \frac{\partial \Phi}{\partial z}\biggr|_{z=z_0} \, z + \,
\frac{1}{2!}\, \frac{\partial^2 \Phi}{\partial z^2}\biggr|_{z=z_0}\, z^2
\nonumber \\ & &  \qquad \qquad + \,
\frac{1}{3!}\, \frac{\partial^3 \Phi}{\partial z^3}\biggr|_{z=z_0}\, z^3 +
\, ...
\,  \nonumber \\
& \equiv & \Phi_0 \,
+ \Phi_{(1)} \, z + \,
\frac{1}{2}\, \Phi_{(2)} \, z^2 + \,
\frac{1}{6}\, \Phi_{(3)} \, z^3 +
\, \cdots
\, .
\label{Phi}
\end{eqnarray}
The definitions of $\Phi_{(j)}\equiv \bigl( {\partial^j \Phi(z)}/{\partial
z^j}\bigr|_{z=z_0} \,= \, - (q E_0)^{(j-1)}_0 + \alpha (B^2)^{(j)}_0 $
(for $j = 1, 2, ...$; the superscript $(j)$ denotes order in partial
differentiation) are obvious
\begin{eqnarray}
\Phi_{(1)} & = & - (q E)_0 + \alpha (B^2)'_0 \, = - q_0 E_0 +
 2 \alpha B_0 B'_0
\nonumber \\
\Phi_{(2)} & = & - (q E_0)'_0 + \alpha (B^2)''_0 \,
\nonumber \\
& & = - (q'_0 E_0 + q_0 E'_0) +
 2 \alpha ({B'}_0^2 + B_0 B''_0)
 \nonumber \\
\Phi_{(3)} & = & - (q E_0)''_0 + \alpha (B^2)'''_0 \,\nonumber \\ & = &
- (q''_0 E_0 + 2 q'_0 E'_0+ q_0 E''_0) +
 2 \alpha (3 B'_0 B''_0 + B_0 B'''_0) \, ,\nonumber \\
&  &
\label{defPhij}
\end{eqnarray}
and so forth.
Note the force balance equation
\[
M g = q_0 E_0 - 2 \alpha B_0 B'_0 \, ,
\]
which is satisfied at equilibrium.

Given the above definitions, the equation of motion (\ref{eqmotion1})
takes the form
\begin{equation}
\ddot{z} + \nu \, \dot{z} \, + \, \omega_g^2 \, z\, +
\, K_1 \, z^2 \, + K_2
\, z^3 \, = \, 0 \, ,
\label{eqmotion2}
\end{equation}
where
\begin{equation}
\omega_g^2 = \Phi_{(2)}/M \, , \quad K_1 = \Phi_{(3)}/(2 M) \, ,
\quad K_2 = \Phi_{(4)}/(6 M) \, , \label{defK12}
\end{equation}
and higher order terms are omitted. One immediately notices the intrinsically nonlinear
character of the transverse dust oscillations due to the electric/magnetic field
inhomogeneity.  In the linear limit ($\Phi_{(j)} = 0$ for $j \ge 3$),
the results of Ref. \cite{Yaro} are exactly recovered.

Once the set of parameter values are determined, Eq. (\ref{eqmotion2}) can
be solved, e.g. via a Lindstedt--Poincar\'e method.
Assuming small displacements of the form
$u(t) = \sum_{n = 1}^\infty \epsilon^n \, u_n(\tau)$,
where the reduced time variable $\tau$ incorporates a
frequency shift due to nonlinearity: $\tau = \omega t =
(1 + \epsilon \lambda_1 + \epsilon^2 \lambda_2 + ...) \omega_g  t \equiv
(1 + \lambda) \omega_g  t$.
One thus obtains the solution
\begin{equation}
u(\tau) \approx \epsilon \, A \, \cos\tau + \,
\epsilon^2 \, \biggl(
\frac{K_1}{3 \omega_g^2} \, A^2 \cos 2 \tau -
 \frac{2 K_1}{\omega_g^2} \, |A|^2
\biggr)+ \cdots \, ,
\label{solution1}
\end{equation}
along with a nonlinear frequency shift $\delta \omega = \lambda
\omega_g \sim |A|^2 \, \omega_g$.  We see that the nonlinearity of
the force acting on the paramagnetic dust grains typically results
in the generation of frequency harmonics once the oscillation
amplitude becomes slightly significant in magnitude. The
occurrence of this effect depends on the relative magnitude of the
coefficients $K_1$ and $K_2$ in Eq. (\ref{eqmotion2}). This
phenomenon has already been studied for an unmagnetized dust
crystal (viz. for $B = 0$) \cite{Ivlev2000}; based on Eq. (9)
therein, one deduces that $\Phi_{(3)}/\Phi_{(2)} = - 1$
mm${}^{-1}$ and $\Phi_{(4)}/\Phi_{(2)} = + 0.42$ mm${}^{-2}$,
implying $K_1/\omega_g^2 = - 0.5$ mm${}^{-1}$ and $K_2/\omega_g^2
= + 0.07$ mm${}^{-2}$.  Similar data on the magnetized DP crystal
case, which are not yet available, may be deduced from appropriate
experiments. This type of analysis may be pursued further once
such feedback from experiments is available.

\section{Nonlinear transverse magnetized dust lattice oscillations}

Let us now consider a layer of identical charged dust grains (of lattice constant
$r_0$).  The Hamiltonian of such a chain is of the form \[H = \sum_n \frac{1}{2} \,
M \, \biggl( \frac{d \mathbf{r}_n}{dt} \biggr)^2 \, + \,
\sum_{m \ne n} U(r_{nm}) \, + \Phi_{ext}(\mathbf{r}_n)\, ,
\]
where $\mathbf{r}_n$ is the position vector of the $n-$th grain.
$U_{nm}(r_{nm})$ is a binary interaction potential function,
related to the electrostatic intergrain interaction potential
$\phi(x)$ (typically of the Debye type, though ion flow in the
sheath may be included for a more complete description
\cite{Ignatov}, as well as the magnetic moment of the $n-$th and
$m-$th grains, which are located at a distance  $r_{nm} =
|\mathbf{r}_{n}-\mathbf{r}_{m}|$. Even though the analytical form
of $U(r)$ need not be specified here, for generality, we may
explicitly refer to the model of Refs. \cite{Yaro, SamsonovNJP} in
the case of magnetized dust crystals [see Eq.(15) in Ref.
\cite{Yaro}, or Eq. (8) in Ref. \cite{SamsonovNJP}].  The external
potential $\Phi_{ext}(\mathbf{r})$ accounts for the external force
fields in the sheath region (i.e. essentially $\Phi$ as defined in
(\ref{Phi}) above); nevertheless, in a more sophisticated
description,  $\Phi_{ext}$ may include the parabolic horizontal
confinement potential imposed in experiments for stability
\cite{Samsonov}, or the initial laser excitation triggering the
oscillations in experiments (both neglected here).

Considering the motion of the $n-$th dust grain in the {\em transverse}
(vertical, off--plane, $\sim \hat z$) direction, we have the equation of
motion including dissipation caused by dust-neutral collisions
\begin{equation}
M\, \biggl( \frac{d^2 z_n}{dt^2} \, + \nu \, \frac{d z_n}{dt}
\biggr) = \,- \sum_n \,\frac{\partial U_{nm}(r_{nm})}{\partial
z_n} \, +
\,
F_{ext}(z_n)
\, , \label{eqmotion0}
\end{equation}
where $F_{ext} = - \partial \Phi_{ext}(z)/\partial z$ accounts for
all external forces in the $z-$ direction.

\subsection{Equation of motion}

Assuming small displacements from equilibrium, one may Taylor
expand the interaction potential $U(r)$ around the equilibrium
intergrain distance $l r_0 =  |n-m| r_0$ (between $l-$th order
neighbors, $l=1, 2, ...$), i.e. around $z_n \approx 0$.
Retaining only nearest-neighbor interactions ($l = 1$), we then obtain
from Eq. (6)
\begin{eqnarray}
\frac{d^2 z_n}{dt^2} \, + \nu \, \frac{d z_n}{dt}
+ \, \omega_g^2 \, z_{n}\,
+ \, K_1 \, z_{n}^2 \,
+ K_2 \, z_{n}^3 \, = \,
\nonumber \\
\omega_{0, T}^2 \, (2 z_{n} - z_{n+1} - z_{n-1}) \,
\, \nonumber \\
+ K_3  \, \biggl[ (z_{n+1} - z_{n})^{3} -
(z_{n} - z_{n-1})^{3} \biggr] \, ,
\label{eqmotion-disc}
\end{eqnarray}
The transverse oscillation characteristic frequency \(\omega_{0,
T}\) and the coefficient $K_3$ are defined by
\begin{eqnarray}
\omega_{0, T} = [- U'(r_0)/(M r_0)]^{1/2} \, , \nonumber \\
K_3 = \, - \frac{1}{2 M r_0^3}\, \bigl[ U'(r_0) - r_0 U''(r_0)
\bigr] \, . \label{defK3}
\end{eqnarray}
The gap frequency $\omega_g$ and the coefficients $K_1$ and $K_2$
have been defined above. Of course, a negative/positive value of
$U'(r_0)$/$U''(r_0)$ is a condition ensuring stability of
transverse dust-lattice (TDL) oscillations, i.e. a real-valued
frequency $\omega_{0, T}$/$\omega_g$ ($\cdot/\dag$ here means
$\cdot$ or $\dag$, respectively); cf. (\ref{defK3}) above.

Notice that the equation of motion presented in Ref. \cite{Yaro}
[see Eq. (18) therein] is exactly recovered in the linear case
(i.e. for $K_1 = K_2 = K_3 = 0$), given the form of the
interaction potential in that model. In particular, the frequency
$\omega_{0, T}$  then becomes \cite{Yaro}
\[
\omega_{0, T}^2 \, = \frac{q^2}{M r_0^3} \, \biggl( 1 + \frac{r_0}{\lambda_D}
\biggr)
\,
\exp\bigl( - \frac{r_0}{\lambda_D}\bigr)
\, + \,
9 \, \frac{m_0^2}{r_0^5}
\, ,
\]
where $\lambda_D$ denotes the Debye length; one immediately
distinguishes the contribution from the Debye potential (first
term) from the second term, which is due to the magnetic moment
$m_0 = \alpha B_0$ (see the definitions above).  Notice, in
passing, that the vertical motion equation of the recent nonlinear
model by Ivlev \textit{et al.} \cite{Ivlev2003} is exactly
recovered in the appropriate limit \cite{note1}.

In general (i.e. regardless of the particular analytical aspects
of each model), one immediately notices that nonlinearity may
either arise from the sheath environment (electric/magnetic
fields) or from the interactions between the paramagnetic dust
grains.

Adopting the standard continuum approximation, often employed in
solid state physics \cite{Kittel}, we may assume that only small
displacement variations occur between neighboring sites, viz. \[
z_{n \pm 1} = u \pm r_0 \frac{\partial u}{\partial x} +
\frac{1}{2} r_0^2 \frac{\partial^2 u}{\partial x^2} \pm \,
\frac{1}{3!} r_0^3 \frac{\partial^3 u}{\partial x^3} +
\frac{1}{4!} r_0^4 \frac{\partial^4 u}{\partial x^4}...\, \],
where the vertical displacement $z_n (t)$ is now expressed by a
continuous function $u = u(x, t)$. One may now proceed by
inserting this ansatz in the discrete equation of motion
(\ref{eqmotion-disc}), and carefully evaluate each term. The
calculation leads to a continuum equation of motion of the form
\begin{eqnarray}
\ddot{u}  \,+ \, \nu \, \dot{u} + c_0^2 \, u_{xx}\,+ c_0^2 \frac{r_0^2}{12} \,
u_{xxxx} \, = \qquad \nonumber \\ \,
\qquad  - \,\omega_g^2 \,
u \,- K_1 u^2 \, - K_2 u^3 + K_3 r_0^4\, (u_x^3)_x
\, , \label{eqmotion-cont}
\end{eqnarray}
where terms of higher nonlinearity have been omitted.
We have defined the characteristic TDLW velocity $c_0 = \omega_{0, T}\, r_0$;
the subscript $x$ denotes
the partial differentiation, viz. $(u_x^3)_x = 3 (u_x)^2 \, u_{xx}$.

The continuum equation of motion (\ref{eqmotion-cont}) is
a modified, damped Boussinesq--like nonlinear equation.
In the following, we omit the phenomenological damping term
(which may be added \textit{a posteriori}, once an explicit solution for the
displacement is obtained).

\section{Amplitude modulation - a Nonlinear Schr\"odinger Equation}

According to the standard reductive perturbation method
\cite{Tsurui}, we shall consider a small displacement of
the form: \( u = \epsilon \, u_1 + \epsilon^2 u_2 +
... \, , \) where $\epsilon \ll 1$ is a small parameter and
solutions $u(x, t)$ at each order are assumed to be a sum of $m-$th
order harmonics, viz. $u_n = \sum_{m=0}^n u_m^{(n)} \, \exp[i (k x
- \omega t)]$ (the reality condition $u_{-m}^{(n)} =
{u_m^{(n)}}^*$ is understood). Time and space scales are
accordingly expanded as
\[
\partial/\partial t \rightarrow \partial/\partial T_0 + \epsilon
\,
\partial/\partial T_1 + \epsilon^2 \partial/\partial T_2 + ...
\]
\[
\equiv \, \partial_0 + \epsilon \,
\partial_1 + \epsilon^2 \partial_2 + ... \, ,
\]
and
\[
\partial/\partial x \rightarrow \partial/\partial X_0 + \epsilon
\,
\partial/\partial X_1 + \epsilon^2 \partial/\partial X_2 + ...
\]
\[
\equiv \, \nabla_0 + \epsilon \, \nabla_1 + \epsilon^2 \nabla_2 +
... \, ,
\]
implying that
\[
\partial^2/\partial t^2 \rightarrow \, \partial_0^2 + 2 \, \epsilon \,
\partial_0 \partial_1 + \epsilon^2 (\partial_1^2 + 2 \partial_0 \partial_2) + ...
\, ,\]
\[
\partial^2/\partial x^2 \rightarrow \, \nabla_0^2 + 2\,  \epsilon \,
\nabla_0 \nabla_1 + \epsilon^2 (\nabla_1^2 + 2 \nabla_0 \nabla_2)
+ ... \, ,
\]
and
\[
\partial^4/\partial x^4 \rightarrow \, \nabla_0^4 + 4\,  \epsilon \,
\nabla_0^3 \nabla_1 + 2 \, \epsilon^2 \nabla_0^2\, (3 \nabla_1^2 +
2 \nabla_0 \nabla_2) + ... \, .
\]
This reductive perturbation technique is a standard procedure, often used in
the study of the nonlinear wave propagation (e.g. in
hydrodynamics, in nonlinear optics, etc.) \cite{Remo}.

We now proceed by substituting all the above series in
(\ref{eqmotion-cont}) and isolating terms arising in the
equation of motion at each order in $\epsilon^n$. By solving for
$u_n$, then substituting in the following order $\epsilon^{n+1}$
and so forth, we obtain the $m-$th harmonic amplitudes
$u_n^{(m)}$ at each order,
along with a compatibility condition up to any given order.

The equation obtained in order $\sim \epsilon^1$ is
\begin{equation}
(\partial_0^2 - c_0^2 \,\nabla_0^2 - c_0^2 \frac{r_0^2}{12} \, \,\nabla_0^4)
\, u_1 \, \equiv \, L_0\, u_1 \,= 0 \, . \label{eq-epsilon1}
\end{equation}
We may assume that $u_1 = u_1^{(1)} \, \exp[i (k x - \omega t)] +
{\rm c.c.} \equiv \exp (i \theta)  + {\rm c.c.}$, where $\omega$, $k = 2
\pi/\lambda$ and $\lambda$ denote the carrier wave frequency,
wavenumber and wavelength, respectively; ${\rm c.c.}$ stands for the
complex conjugate.
The dispersion relation obtained from (\ref{eq-epsilon1}) is of
the form
\begin{eqnarray}
\omega^2 & = & \omega_g^2 \, - k^2 \,c_0^2 \,  + k^4 \, \frac{c_0^2}{12} \, r_0^2 \,
\nonumber \\
& = & \,
\omega_g^2 \, - \omega_{T, 0}^2 \, k^2 \, r_0^2\,
\biggl(1 - \, \frac{1}{12} k^2 \, r_0^2 \biggr) \, ,
\label{dispersion}
\end{eqnarray}
predicting an inverse optical-mode-like behaviour (in agreement
with previous results; in fact, this is exactly the continuum
analogue of the discrete dispersion relation in Ref. \cite{Yaro}).
We assume that the reality condition $\omega > 0$ is, in
principle, satisfied in the region of validity of the continuum
hypothesis $k \ll (\pi/r_0) \equiv k_{cr, 0}$ (nevertheless, note
that the possibility of magnetic field--related instabilities,
basically related to the relative magnitude of $\omega_g^2$ and
$\omega_{T, 0}^2$, was put forward in Ref. \cite{Yaro}).

Let us evaluate the action of the linear operator $L_0$, defined
above, on higher harmonics of the phase $\theta$
\begin{eqnarray}
L_0\, e^{i n \theta} &=& [(-i n \omega)^2 + \omega_g^2 + c_0^2 (i n k)^2 + c_0^2
\, (r_0^2/12) (i n k)^4 ] \,\nonumber \\
&& \qquad \times e^{i n \theta} \nonumber \\
& =& \bigl[ n^2 (n^2 - 1)
c_0^2 \, (r_0^2/12) \, k^4 - (n^2 - 1) \omega_g^2 \bigr] \,e^{i n \theta}
\nonumber \\
&\equiv & D_n \, e^{i n
\theta}\, , \label{Dn}
\end{eqnarray}
where we made use of the dispersion relation (\ref{dispersion});
thus $D_0 = \omega_g^2$, $D_1 = 0$ and $D_2 = - 3 \omega_g^2 +
c_0^2 r_0^2 k^4$.

To order $\sim \epsilon^2$, the condition of suppression of secular terms
takes the form
\begin{equation}
\frac{\partial u_1^{(1)}}{\partial T_1 } + v_g\, \frac{\partial
u_1^{(1)}}{\partial X_1 } = 0\, ,  \label{secular1}
\end{equation}
i.e. $u_1^{(1)} = u_1^{(1)}(\zeta)$, where $\zeta = \epsilon (x -
v_g t)$, implying that the slowly--varying wave envelope travels
at the {\em{negative}} \textit{group velocity} $v_g = \omega'(k) =
- (1 - k^2 r_0^2 /6) \, c_0^2 \,k/\omega$; this {\em{backward}}
wave has been observed experimentally \cite{Misawa}. We notice
that $v_g$ becomes zero at $k = \sqrt{6} \,r_0^{-1}$, beyond which
it becomes positive; nevertheless, note for rigor that this
comment rather becomes obsolete once one properly takes into
account the influence of the lattice discreteness on the form of
the dispersion relation $\omega = \omega(k)$; cf. Ref.
\cite{IKPKSPoP2004}.

This procedure finally leads to a solution of the form
\begin{eqnarray}
u(x, t) & = & \epsilon \,\biggl[A \,e^{i \,(k x - \omega t)} + c.c.\biggr] \
\nonumber \\
&& + \, \epsilon^2 \,K_1\, \biggl[ - \frac{2 \, |A|^2}{\omega_g^2}
\, + \frac{A^2}{-D_2} \,e^{2 i \,(k x - \omega t)} + c.c.
\biggr] \, \nonumber \\
&&  + {\cal O}(\epsilon^3) \, .
\end{eqnarray}
where $\omega$ obeys the dispersion relation (\ref{dispersion}).

The amplitude $A(X, T)$ obeys a {\em Nonlinear Schr\"{o}dinger Equation}
(NLSE) of the form
\begin{equation}
i\, \frac{\partial A}{\partial T} + P\, \frac{\partial^2
A}{\partial X^2} + Q \, |A|^2\,A = 0 \, , \label{NLSE}
\end{equation}
where the `slow' variables $\{ X, T \}$ are $\{ X_1 - v_g \, T_1,
T_2 \}$, respectively. The {\em dispersion coefficient} $P$, which
is related to the curvature of the phonon dispersion relation
(\ref{dispersion}) as $P = \,(1/2) ({d^2 \omega}/{d k^2})$, reads
\begin{equation}
P =  - \frac{c_0^2 \omega_g^2}{2 \omega^3}\biggl( 1 -  \frac{1}{2}\, k^2 r_0^2
\biggr) \, , \label{Pcoeff}
\end{equation}
and the {\em nonlinearity coefficient}
\begin{equation}
Q = \frac{1}{2 \omega} \biggl[ 2 K_1^2 \biggl(
\frac{2}{\omega_g^2} +  \frac{1}{D_2} \biggr)
 - 3\, K_2 - 3 K_3 \,k^4 r_0^4 \biggl] \,
\label{Qcoeff}
\end{equation}
is related to both the sheath electric/magnetic field and the
intergrain coupling nonlinearity discussed above. As expected,
(\ref{Qcoeff}) recovers exactly the previously derived expressions
(in similar studies) in the appropriate limits, i.e. precisely Eq.
(9) in \cite{IKEPS29} in the dispersionless  limit (for very low
$k$, i.e. cancelling terms in $k^4$ in (\ref{dispersion})), and
Eq. (10) in \cite{IKPKSPoP2004} (where a quasi-continuum
approximation was adopted, i.e. a continuum envelope but discrete
carrier wave desciption). Notice however that inter-grain
interaction nonlinearity was neglected in those studies (i.e. $K_3
= 0$ therein).

In view of the analysis that follows, it may be appropriate to
cast Eq. (\ref{NLSE}) in a non-dimensional form, for physical
clarity. This is readily done by introducing a set of
appropriately chosen scales, e.g. the lattice constant $r_0$ and
the inverse eigenfrequency $\omega_g^{-1}$, i.e. by scaling the
(slow) variables $X$ and $T$ as $X \rightarrow X' =  X/r_0$ and $T
\rightarrow T' = \omega_g T$; the vertical displacement amplitude
becomes $A \rightarrow A' = A/r_0$. The form of Eq. (\ref{NLSE})
is then exactly recovered upon substituting with: Eq. $P
\rightarrow P' = P/(\omega_g r_0^2)$ and $Q \rightarrow Q' = Q
r_0^2/\omega_g$. The primes will be dropped in the following.

\paragraph{Modulational instability.}

In a generic manner, a modulated wave whose amplitude obeys the
NLS equation (\ref{NLSE}) is stable/unstable to perturbations if
the product \( P Q \) is negative/positive.
To see this, one may first check that the NLSE is satisfied by the
monochromatic solution (Stokes' wave)
\( A(X, T) = A_0\, e^{i Q
|A_0|^2 T} \, + \, c.c. \) The standard (linear) stability
analysis then shows that a linear modulation with the frequency
$\Omega$ and the wavenumber $\kappa$ obeys the dispersion relation
\begin{equation}
\Omega^2(\kappa) = P^2\, \kappa^2\, \biggl( \kappa^2\, - 2
\frac{Q}{P}\,|A_0|^2 \biggr) \, ,
\label{pert-disp}
\end{equation}
which exhibits a purely growing mode if
\(\kappa \geq \kappa_{cr} =
({Q}/{P})^{1/2}\,|A_0|\). The growth rate attains a maximum value
\( \gamma_{max} =
{Q}\,|A_0|^2 \). This mechanism is known as
the {\sl Benjamin-Feir instability} \cite{Remo}.
For $P Q < 0$, the wave is modulationally stable,
as evident from (\ref{pert-disp}).

Notice that the coefficient $P$ is negative, given the
quasi-inverse-parabolic form of the dispersion curve $\omega(k)$
for low $k$. One therefore only needs to deduce the sign of Q,
given by (\ref{Qcoeff}), in order to determine the stability
profile of the TDL oscillations. In fact, given the definitions
(\ref{defK12}, \ref{defK3}) of the parameters $\omega_g$, $K_1$,
$K_2$ and $K_3$, one may easily derive an expression for $Q$ in
terms of the (derivatives of the) potentials $\Phi(z)$ (for the
sheath) and $U(z)$ (for inter-grain interactions). The exact form
of the potential $\Phi(z)$ (hence the coefficients $\Phi_{(j)}$,
$j = 1, 2, ...$) may be obtained from \textsl{ab initio}
calculations or by experimental data fitting. As mentioned above,
some evidence for the numerical values of the coefficients
$\Phi_{(j)}$, yet only in the absence of the magnetic field, can
be found in Ref. \cite{Ivlev2000}, where the dust grain potential
energy was reconstructed from experimental data (see Eq. (9)
therein); those values (cf. comment above) seem to suggest that
the value of $Q$ for low wavenumbers $k$ is positive, as may be
readily checked from (\ref{Qcoeff}). Therefore, under the
experimental conditions described in Ref. \cite{Ivlev2000}, the
TDL wave would propagate as a \emph{stable} wave, for large
wavelength values $\lambda$. However, for shorter wavelengths,
either the coefficient $P = \, \omega''(k)/2$ or $Q$ may change
sign, and the TDL wave may thus be potentially unstable. These
results should \textsl{a priori} be checked by appropriately
designed experiments, in particular with regard to magnetically
levitated DP crystals.

\paragraph{Localized excitations.}

A final comment concerns the possibility of the existence of
localized excitations related to TMDL waves. It is known that the
NLSE (\ref{NLSE}) admits localized solutions (envelope solitons)
of the {\sl bright} ($P Q > 0$) or {\sl dark/grey} ($P Q < 0$)
type. These expressions are found by inserting the trial function
\( A = A_0 \exp(i\Theta) \) in Eq. (\ref{NLSE}) and then
separating real and imaginary parts in order to determine the
(real) functions $A_0(X, T)$ and $\Theta(X, T)$. Details on the
derivation of their analytic form can be found e.g. in Refs.
\cite{Fedele, Fedele2}, so only the final expressions will be
given in the following. Let us retain that this \textit{ansatz}
amounts to a total (vertical) grain displacement $u(x, t)$
essentially equal to:
\begin{equation}
u(x, t) = \epsilon \rho \, \cos(k x - \omega t + \Theta) \, ,
\label{totalsolution}
\end{equation}
where $\rho = 2 A_0$ and the nonlinear phase-shift is $\Theta \sim
\epsilon$, since linear in $\{X, T\} = \{ \epsilon  x, \epsilon
t\}$ (see below).

The \emph{bright} type ({\em{pulse}}) envelope solutions (\textsl{continuum
breathers}, see Figs. \ref{fig1}, \ref{fig2}), obtained for $P Q > 0$, 
are given by
\begin{eqnarray}
A &=& \biggl( \frac{2 P}{Q L^2} \biggr)^{1/2} \, {\rm{sech}}
\biggl( \frac{X - v_e \,
T}{L} \biggr)\, \nonumber \\
&& \qquad \times \exp \biggl\{ i \, \frac{1}{2 P} \bigl[ v_e  X +
\bigl(\Omega - \frac{v_e^2}{2} \bigr) T \bigr] \biggr\}\, ,\label{breather}
\end{eqnarray}
where $v_e$ is the envelope velocity; $L$ and $\Omega$ represent
the pulse's spatial width and oscillation frequency, respectively.
In our problem, the bright-type localized envelope solutions may
occur and propagate in the lattice if a sufficiently short
wavelength is chosen, so that the product $P Q$ is positive. We
note that the pulse width $L$ and the amplitude $A_0$ satisfy $L
A_0 = (2 P/Q)^{1/2} = const.$. 
Let us point out that, when the pulse's spatial width $L$ is
comparable in order of magnitude to the carrier wavelength $\lambda$, 
this (now highly localized) type of solution  
is similar in structure to
the (odd-parity) discrete breather (DB) modes (or intrinsic localized modes, 
ILMs) recently widely studied in molecular
chains \cite{breathers}. 

For $P Q < 0$, we have the {\em dark} envelope
soliton ({\em{hole}}) \cite{Fedele} (see Fig. \ref{fig3})
\begin{eqnarray}
A &=& \pm A_0 \, {\rm{tanh}}
\biggl( \frac{X - v_e \,
T}{L'} \biggr)\, \nonumber \\
&& \qquad \times \exp \biggl\{ i \, \frac{1}{2 P} \bigl[ v_e  X + 
\bigl( 2 P Q A_0^2 - \frac{v_e^2}{2} \bigr) T \bigr] \biggr\}\, ,
\label{darksoliton}
\end{eqnarray}
which represents a localized region of negative wave density (void). 
The pulse width $L'
= (2 |P/Q|)^{1/2}/A_0$ is inversely proportional to the
amplitude $A_0$. 

For $P Q < 0$, we also have the {\em grey} envelope
soliton \cite{Fedele} (see Fig. \ref{fig4})
\begin{equation}
A \, = \, A_0 \, \{ 1 - d^2 \,  {\rm{sech}}^2\{[X - v_e \,
T]/L''\}\}^{1/2} \, \exp ( i\, \Theta ) \, , \label{greysoliton}
\end{equation}
where $\Theta =  \Theta(X, T)$ is a nonlinear phase correction to
be determined by substituting into the NLSE (\ref{NLSE}); the
calculation yields the complex expression:
\begin{eqnarray}
\Theta & = & \frac{1}{2 P} \, \biggl[ V_0\,X \, -
\biggl(\frac{1}{2} V_0^2 - 2 P Q A_0 \biggr) \,T + \Theta_{0}
\biggr]\, \nonumber \\ &  & \qquad \, - S \, \sin^{-1} \frac{d\,
\tanh\bigl(\frac{X - v_e\, T}{L''} \bigr)}{\biggr[  1 - d^2\,
{\rm{sech}}^2 \biggl(\frac{X - v_e\, T}{L''} \biggr) \biggr]^{1/2}}
\, , \label{greysoliton-Theta}
\end{eqnarray}
(see (64) in Ref. \cite{Fedele}a). 
This localized excitation 
also represents a localized region of negative wave density;
$\Theta_{0}$ is a constant phase; $S$ denotes the product $S = 
{\rm{sign}} (P) \, \times {\rm{sign}} (v_e - V_0)$. Again, the pulse width $L'
= (|P/Q|)^{1/2}/(d A_0)$ is inversely proportional to the
amplitude $A_0$, and now also depends on an independent real parameter
$d$, which regulates the modulation depth; $d$ is given by: \( d^2
\, = \, 1 \, + \, (v_e^2 - V_0^2)/({2 P Q} {A_0}) \, \le \, 1 \). 
$V_0$ is an independent real constant
which satisfies (see details in Ref. \cite{Fedele}): \( V_0 -
\sqrt{2 |P Q|\, A_0} \, \le \, v_e \, \le \,V_0 + \sqrt{2 |P Q|\,
A_0} \). This excitation represents a localized region of negative
wave density (a \emph{void}), with finite amplitude 
at $X = 0$. For $d = 1$ (thus $V_0 = v_e$), one recovers the {\em
dark} envelope soliton presented above, which is characterized by a vanishing
amplitude at $X = 0$. 

We should admit, for rigor, that the latter
excitations (of dark/grey type), yet apparently privileged in the
continuum limit (where $P Q < 0$ for low $k$), are rather
physically irrelevant in our (infinite chain) model, since they
correspond to an infinite energy stored in the lattice.
Nevertheless, their existence locally in a finite--sized chain may
be considered (and possibly confirmed) either numerically or
experimentally.

It may be stressed that the grain displacement corresponding to 
the (envelope) excitations presented here is 
intrinsically different in form (and obeys different physics) from
the pulse-like small-amplitude localized structures (solitons)
typically found via Korteweg-DeVries (KdV) theories; see Refs.
\cite{Fedele, Fedele2} for a critical comparison. 

\section{Conclusions}

The present study has been devoted to an investigation of
amplitude modulation effects associated with either isolated
transverse dust-grain oscillations or propagating transverse
dust--lattice waves in DP crystals of paramagnetic charged dust
particles, embedded in an external magnetic field. We have shown
that nonlinearity comes into play once the oscillation amplitude
slightly departs from the weak--displacement (linear) regime. This
nonlinearity, which is related to the electric and/magnetic
field(s) in the sheath region, affects the dynamics of transverse
dust lattice oscillations via the generation of phase harmonics
and the potential instability of the carrier wave, due to
self-interaction.  The latter may presumably be responsible for
energy localization in the DP crystal via the formation of
localized envelope excitations.  Analytic expressions for these
excitations are presented and briefly discussed.  It should be
stressed that the instability suggested here, related to
self-modulation of the carrier wave and triggered once the
amplitude becomes slightly important, is completely independent
from the one pointed out in Ref. \cite{Yaro}, which is related to
the values of the intrinsic parameters (the gap frequency, in
particular) involved in the physics of the problem.

\medskip

\begin{acknowledgments}
This work was supported by the European Commission (Brussels)
through the Human Potential Research and Training Network via the
project entitled: ``Complex Plasmas: The Science of Laboratory
Colloidal Plasmas and Mesospheric Charged Aerosols'' (Contract No.
HPRN-CT-2000-00140).
\end{acknowledgments}

\newpage


\centerline{\textbf{Figure Captions}}

Figure 1.

\emph{Bright} type (pulse) soliton solution ($P Q > 0$) of the NLS
equation (\ref{NLSE}), obtained for an indicative set of numerical
values for the parameters in Eqs. (\ref{totalsolution},
\ref{breather}): $\epsilon= 0.1$, $P = Q = 1$, $v_e = 0.2$
(i.e. $\rho = 2 A_0 = \sqrt{2}$), 
$k = 2 \pi/\lambda = 2 \pi$ and $L = 2$: 
(a) the waveform, as results from
(\ref{totalsolution}); (b) the corresponding (localized) wave
energy ($\sim \rho^2$) -- normalized value.

\vskip 1cm

Figure 2.

\begin{figure}[htb]
 \centering
 \resizebox{3in}{!}{
 \includegraphics[]{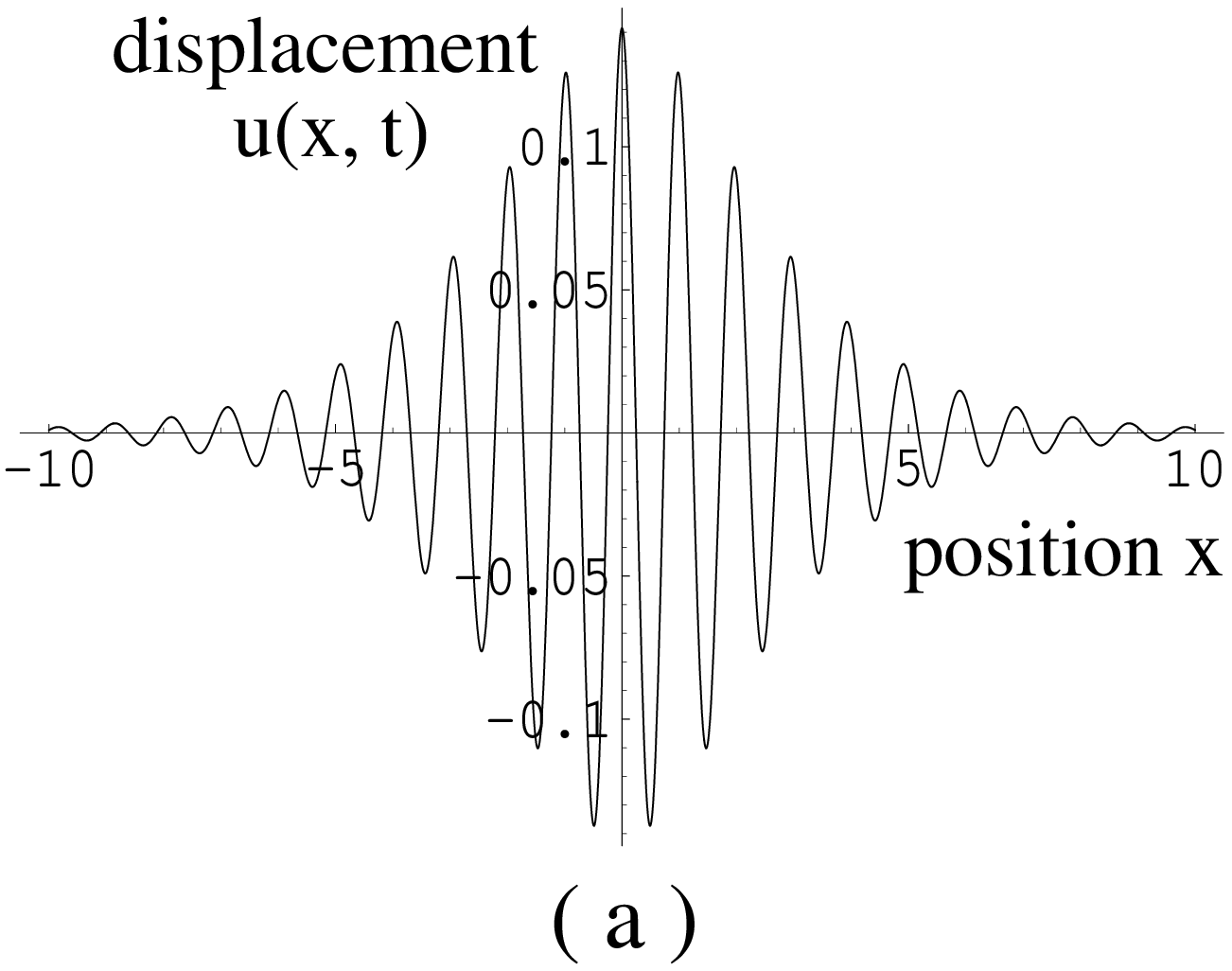}}
\\
\vskip 2 cm \resizebox{3in}{!}{
\includegraphics{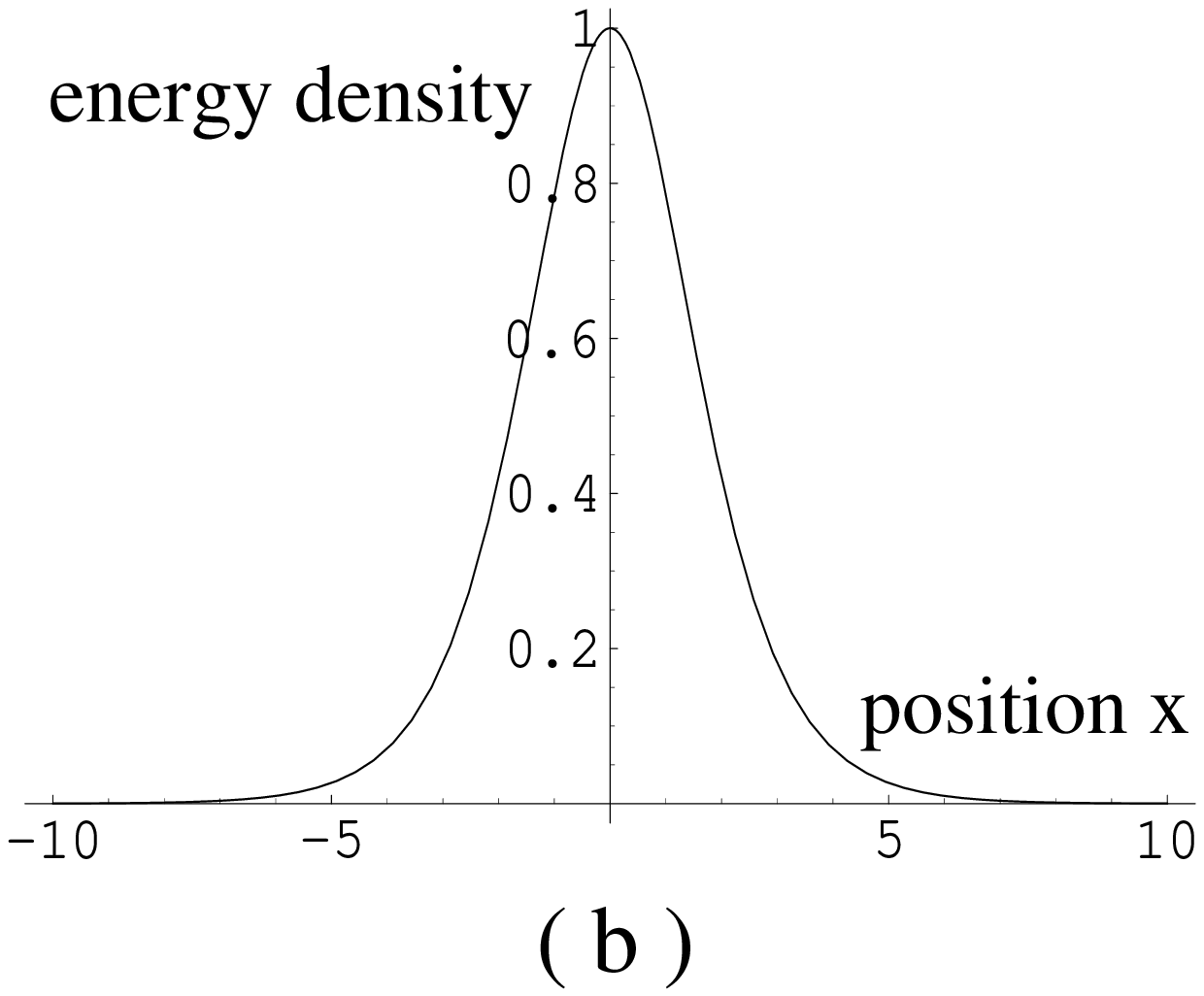}
} \caption{} \label{fig1}
\end{figure}

\begin{figure}[htb]
 \centering
 \resizebox{3in}{!}{
 \includegraphics[]{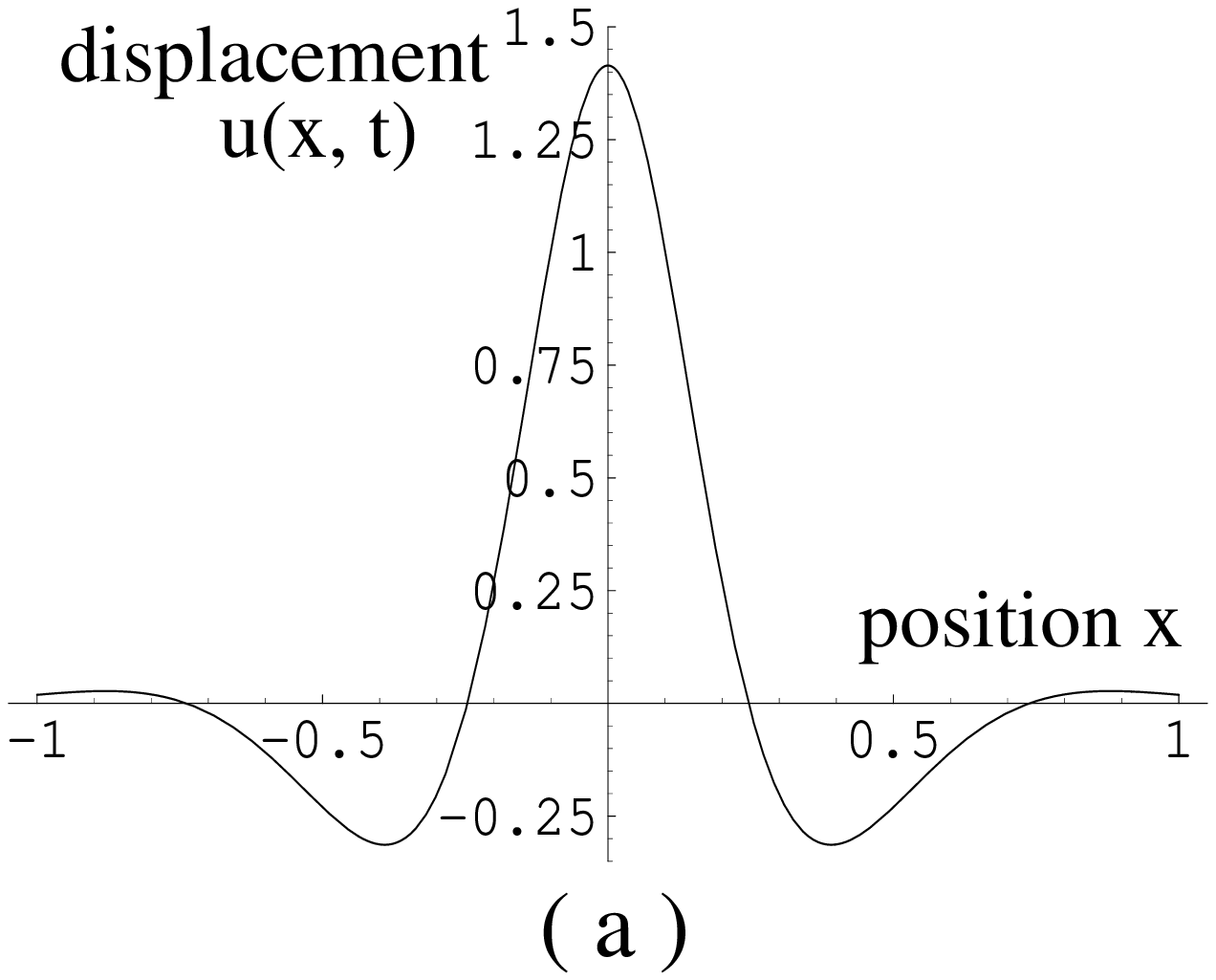}}
\\
\vskip 2 cm \resizebox{3in}{!}{
\includegraphics{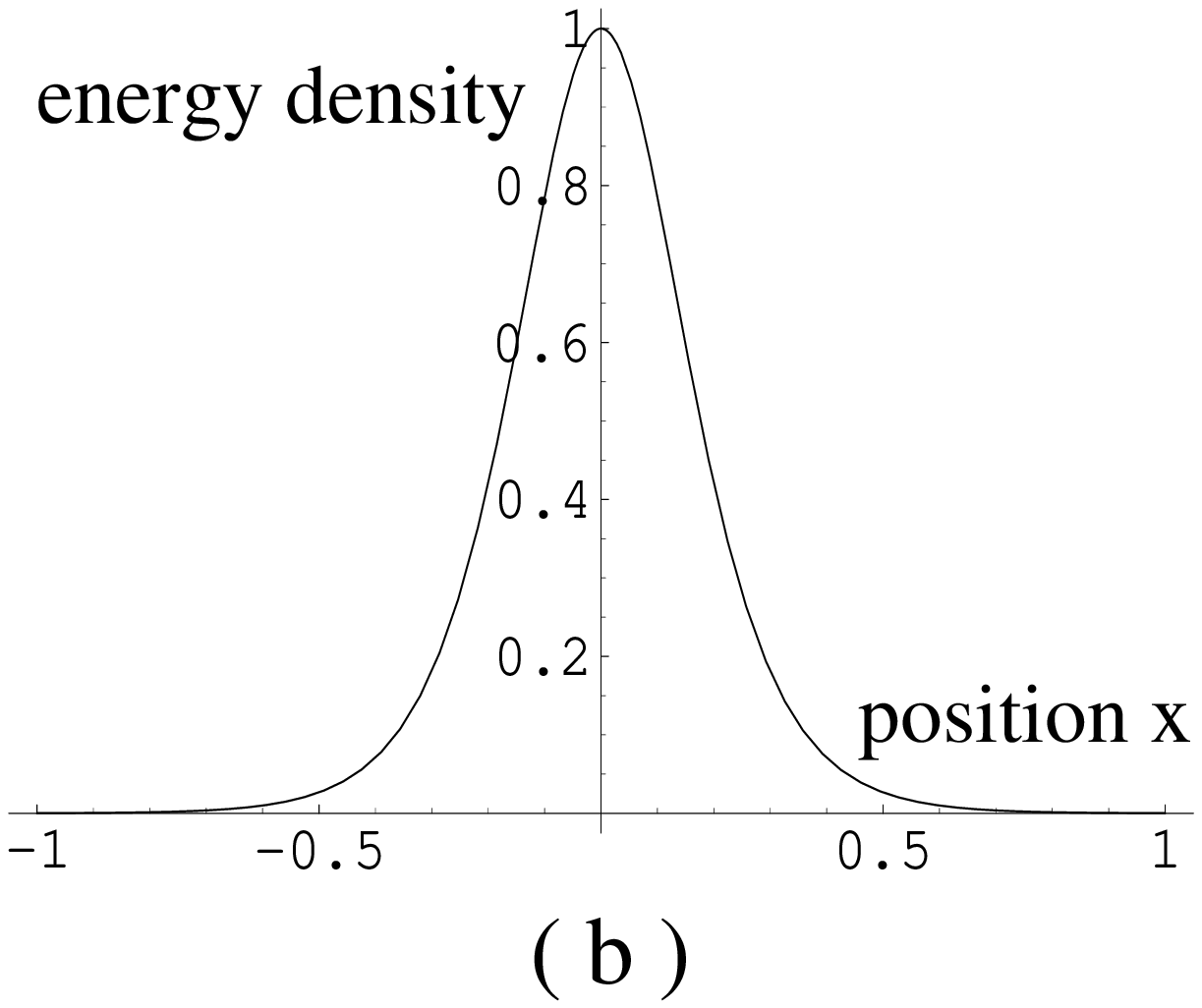}
} \caption{} \label{fig2}
\end{figure}

Same label and data as in Fig. \ref{fig1}, except $L = 0.2$. 

\vskip 1cm

Figure 3.

\emph{Dark} type soliton solutions of the NLS equation for $P Q <
0$ (holes); here $Q = - P = 1$ and the remaining values are just as in 
Fig. \ref{fig1}:
(a) the waveform, as results from
(\ref{totalsolution}); (b) the corresponding (localized) wave
energy ($\sim \rho^2$), normalized over it asymptotic value.

\vskip 1cm

Figure 4.

\begin{figure}[htb]
 \centering
 \resizebox{3in}{!}{
 \includegraphics[]{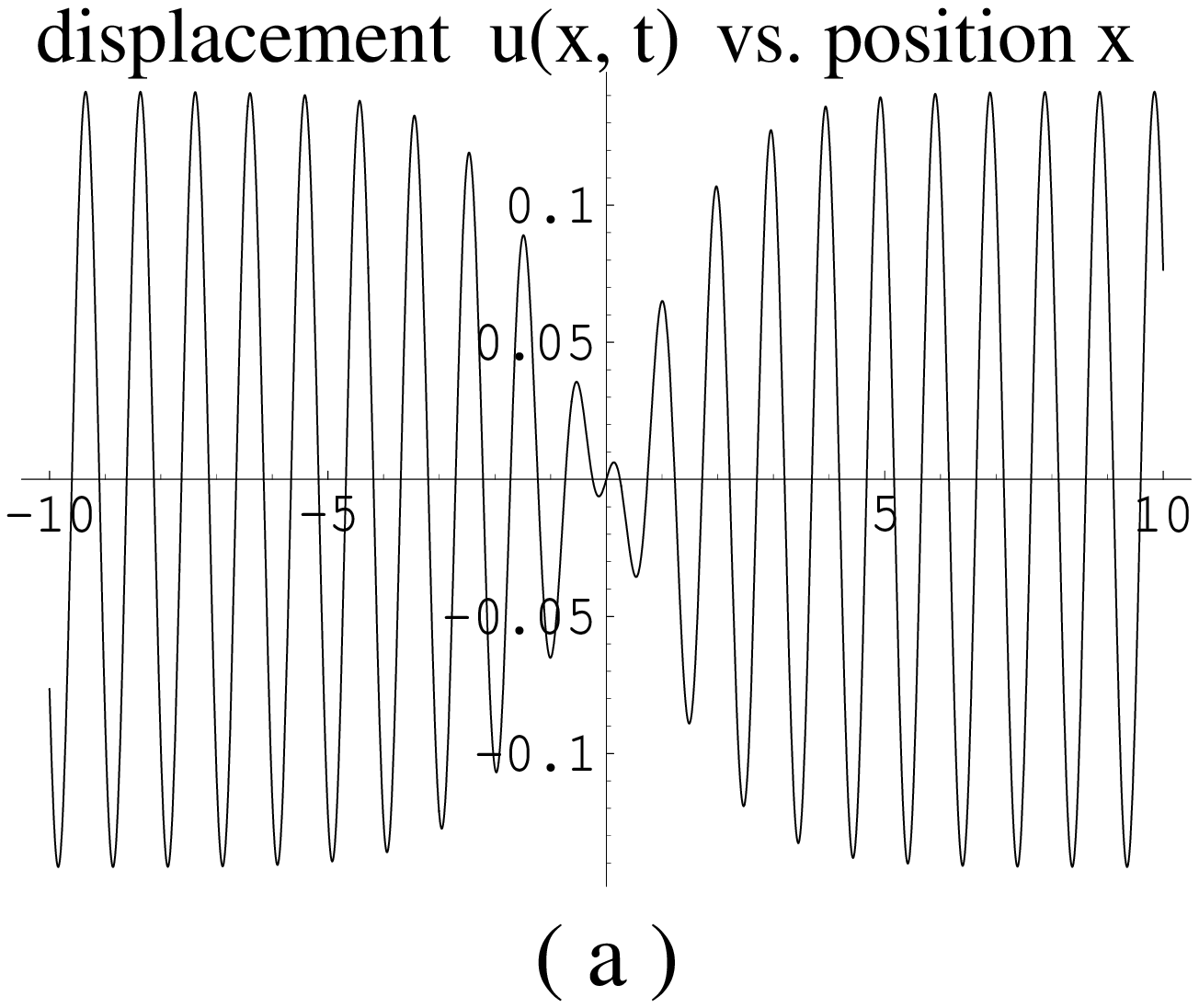}}
\\
\vskip 2 cm \resizebox{3in}{!}{
\includegraphics{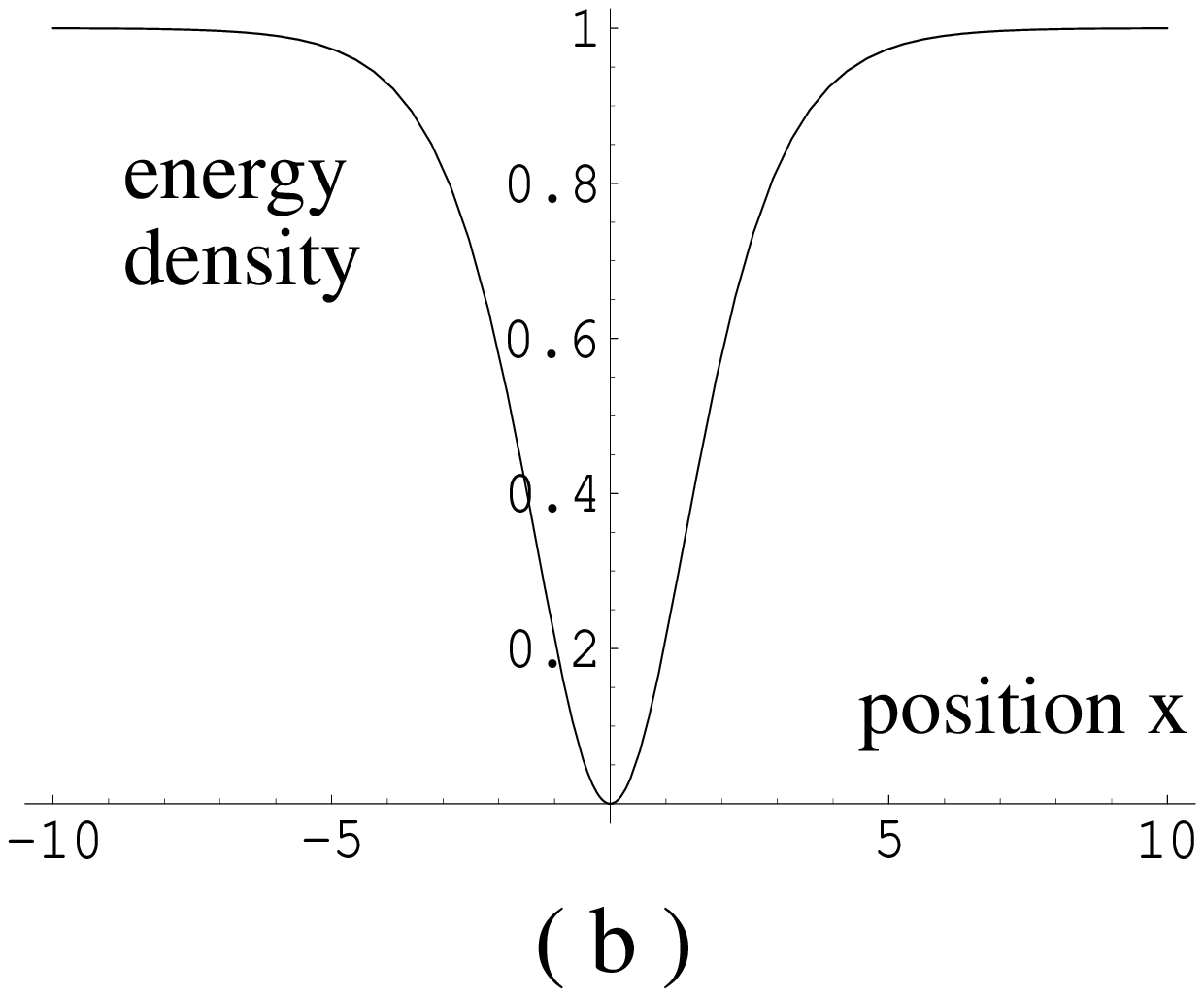}
} \caption{} \label{fig3}
\end{figure}

\begin{figure}[htb]
 \centering
 \resizebox{3in}{!}{
 \includegraphics[]{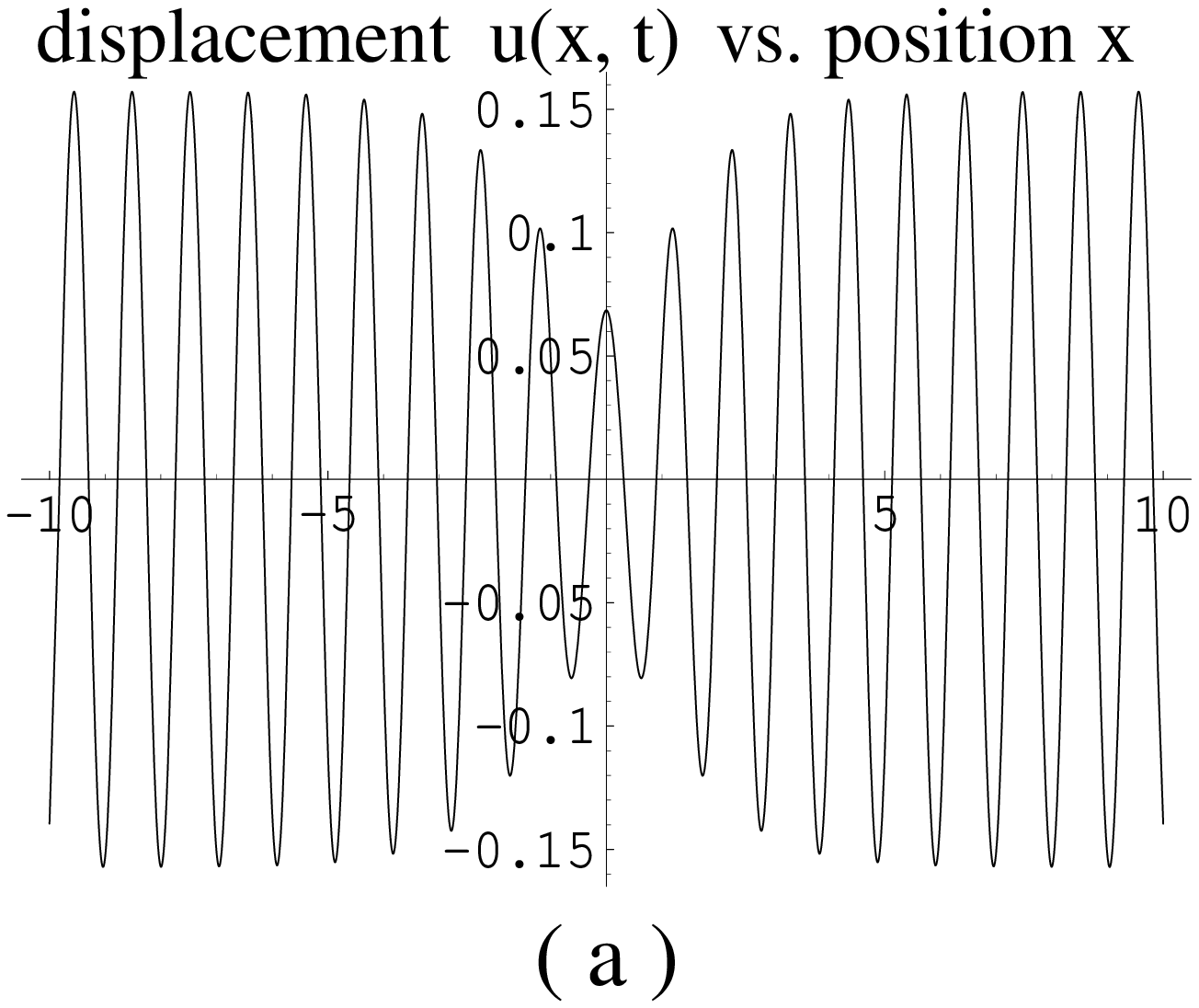}}
\\
\vskip 2 cm \resizebox{3in}{!}{
\includegraphics{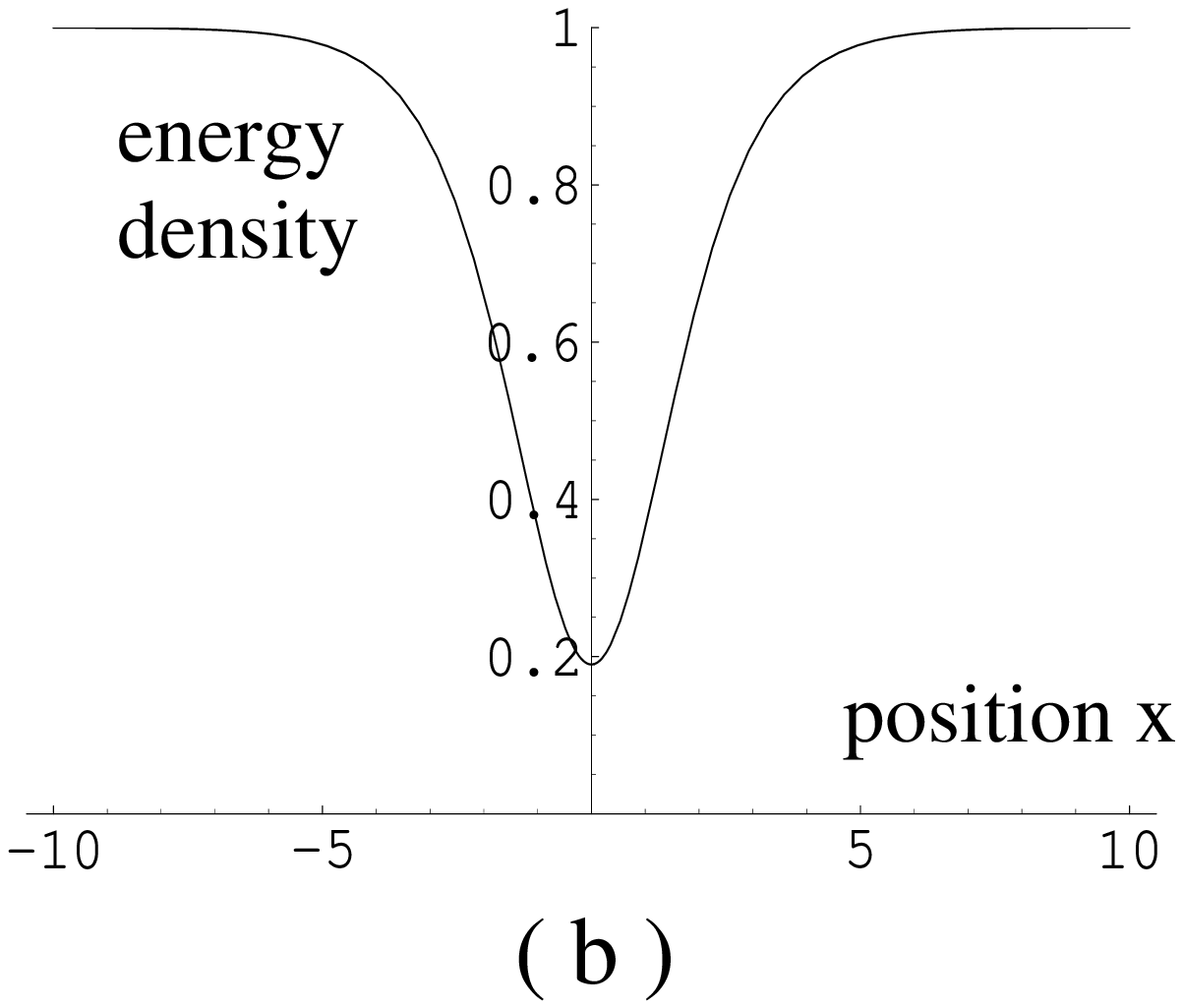}
} \caption{} \label{fig4}
\end{figure}

\emph{Grey} type soliton solutions of the NLS equation for $P Q <
0$: values identical to those in Fig. \ref{fig3}, in addition to 
$V_0 = 0.5$, $d= 0.9$. 
Notice that the amplitude never reaches zero.


\newpage

\newpage

\newpage

\newpage

\newpage

\begin{thebibliography}{10}

\bibitem{PKSbook} See e.g. in: P. K. Shukla and A. A. Mamun,
\textit{Introduction to Dusty Plasma Physics} (Institute of Physics, Bristol, 2002);
also References therein.

\bibitem{Morfill} G. E. Morfill, H. M. Thomas and M. Zuzic, in
 \textit{Advances in Dusty Plasma Physics}, Eds. P. K. Shukla,
 D. A. Mendis and T. Desai (Singapore, World Scientific) p. 99.

\bibitem{SamsonovNJP} D. Samsonov, S. Zhdanov, G. Morfill and V. Steinberg,
New J. Phys. \textbf{5}, 24.1 (2003).

\bibitem{YaroNJP} V. V. Yaroshenko, G. E. Morfill, D. Samsonov and
S. V. Vladimirov, New J. Phys. \textbf{5}, 18.1 (2003).

\bibitem{Yaro} V. V. Yaroshenko, G. E. Morfill and D. Samsonov,
Phys. Rev. E \textbf{69}, 016410 (2004).

\bibitem{Ivlev2000PRE622739} A. V. Ivlev, U. Konopka, and G. Morfill,
Phys. Rev. E \textbf{62}, 2739 (2000).

\bibitem{Jackson} J. D. Jackson, \textit{Classical Electrodynamics}
(John Wiley and Sons, New York, 1963).

\bibitem{Ivlev2000} A. V. Ivlev, R. S\"utterlin, V. Steinberg, M. Zuzic
and G. Morfill, Phys. Rev. Lett. \textbf{85}, 4060 (2000).

\bibitem{Ignatov}
A. M. Ignatov, Plasma Physics Reports {\bf 29}, 296 (2003);
I. Kourakis \, and P. K. Shukla, Phys. Lett. A \textbf{317}, 156 (2003).

\bibitem{Samsonov} D. Samsonov, A. V. Ivlev, R. A. Quinn, G. Morfill  and
S. Zhdanov, Phys. Rev. Lett. \textbf{88}, 095004 (2002).

\bibitem{Ivlev2003} A. V. Ivlev, S. K. Zhdanov, and G. E. Morfill
Phys. Rev. E \textbf{68}, 066402 (2003).

\bibitem{note1} i.e. ignoring $K_1$, $K_2$ and $m_0$;
check by setting $\delta x_{\pm} = 0$ in Eq. (2) in Ref. \cite{Ivlev2003}.

\bibitem{Kittel} C. Kittel,
\textit{Introduction to Solid State Physics} (John Wiley and Sons,
New York, 1996).

\bibitem{Tsurui} A. Tsurui, Progr. Theor. Phys. {\bf 48}, 1196
(1972).

\bibitem{Remo} M. Remoissenet,
\textit{Waves Called Solitons} (Springer-Verlag, Berlin, 2nd Ed.,
1996).

\bibitem{Misawa}
T. Misawa, N. Ohno, K. Asano, M. Sawai, S. Takamura and P. K. Kaw,
Phys. Rev. Lett. {\bf 86}, 1219 (2001).

\bibitem{IKPKSPoP2004} I. Kourakis \, and P. Shukla,
\textit{Nonlinear modulation of transverse waves in dusty plasma
crystals}, Phys. Plasmas \textbf{11}, in press (sch. 05/2004);
also at: {\tt{http://arxiv.org/abs/cond-mat/0402261}}.


\bibitem{IKEPS29} I. Kourakis, {\it Proceedings of the
29th EPS meeting on Controlled Fusion and Plasma Physics}, {\it
European Conference Abstracts (ECA)} Vol. 26B P-4.221 (European
Physical Society, Petit-Lancy, Switzerland, 2002).


\bibitem{Fedele} R. Fedele and H.
Schamel, {\it Eur. Phys. J. B} {\bf 27} 313 (2002); R. Fedele, H. Schamel and P. K. Shukla,
{\it Phys. Scripta} T {\bf 98} 18 (2002).

\bibitem{Fedele2} R. Fedele, {\it Phys.
Scripta} {\bf 65} (6) 502 (2002).

\bibitem{breathers}
G. Tsironis and E. N. Economou (Eds.), \textit{Fluctuations,
Disorder and Nonlinearity}, Physica D \textbf{113} (North-Holland,
Amsterdam, 1998): see several papers in this volume; also, T.
Bountis, H.W. Capel, M. Kollman, J.M. Bergamin, J.C. Ross and J.P.
van der Weele,, Phys. Lett. {\bf 268}, 50 (2000).

\end{thebibliography}
\end{document}